\documentclass[a4paper,fleqn,usenatbib]{mnras}

\usepackage[T1]{fontenc}
\usepackage{ae,aecompl}

\usepackage{graphicx}	
\usepackage{amsmath}	
\usepackage{amssymb}	

\newcommand{\hu}{H\'enon units}
\newcommand{\nbody}{$N$-body\ }
\newcommand\ltorder{\mathrel{\raise.3ex\hbox{$<$}\mkern-14mu
             \lower0.6ex\hbox{$\sim$}}}
\newcommand\comment[1]{{#1}}

\title[Weinberg's mode in an \nbody system]{$l=1$: Weinberg's weakly damped mode in an $N$-body model of a spherical stellar system}

\author[D.C. Heggie et al.]{
Douglas C. Heggie,$^{1}$\thanks{E-mail: d.c.heggie@ed.ac.uk}
Philip G. Breen,$^{1}$
Anna Lisa Varri$^{2,3}$
\\
$^{1}$School of Mathematics and Maxwell Institute for Mathematical Sciences, University of Edinburgh, Kings Buildings, Edinburgh EH9 3FD,UK\\
$^{2}$Department of Astronomy, School of Science,
The University of Tokyo,
7-3-1 Hongo, Bunkyo-ku,
Tokyo, 113-0033
Japan\\
$^{3}$Institute for Astronomy, University of Edinburgh, Royal Observatory,
Blackford Hill, Edinburgh, EH9 3HJ, UK
}

\date{Accepted 2020 January 3. Received 2019 November 23; in original form 2019
April 7}

\pubyear{2020}

\begin{document}
\label{firstpage}
\pagerange{\pageref{firstpage}--\pageref{lastpage}}
\maketitle

\begin{abstract}
  Spherical stellar systems such as King models, in which the distribution function is a decreasing function of energy and depends on no other invariant, are stable in the sense of collisionless dynamics.  But Weinberg 
  showed, by a clever application of the matrix method of linear stability, that they may be nearly unstable, in the sense of possessing {\sl weakly} damped modes of oscillation.  He also demonstrated the presence of such a mode in an $N$-body model by endowing it with initial conditions generated from his perturbative solution.  In the present paper we \comment{provide evidence for} 
  the presence of this same mode in $N$-body simulations of the King $W_0 = 5$ model, in which the initial conditions are generated by the usual Monte Carlo sampling of the King distribution function.  It is shown that the oscillation of the density centre correlates with variations in the structure of the system out to a radius of about 1 virial radius, but anticorrelates with variations beyond that radius.  Though the oscillations appear to be continually reexcited (presumably by the motions of the particles) we show by calculation of power spectra that \comment{Weinberg's estimate of the period}  (strictly, \comment{$2\pi$ divided by} the real part of the eigenfrequency)
  \comment{lies within the range where the power is largest.}  
  \comment{In addition, however, the power spectrum displays another very prominent feature at shorter periods, around 5 crossing times.}
\end{abstract}




\section{Introduction}

There is no shortage of research on the stability of spherical stellar systems, and it uses three basic approaches.  First are rigorous results (theorems, really) based on variational techniques and allied methods \citep[][]{
  1971PhRvL..26..725D,
1980CRASB.290..545G,
1987IAUS..127..531A,
1996MNRAS.280..689P}.
Second are numerical methods based on the equations of motion\footnote{We also include \citet{1979MNRAS.189..723S}, who used numerical integration of the collisionless Boltzmann equation.}, such as tree codes \citep{
1973AA....24..229H,
1988ApJ...328...93D,
1996MNRAS.280..700P,
1999MNRAS.305..859A,
  2002AA...395...25M}.  Useful introductions to these two kinds of technique are provided by \citet{2008gady.book.....B}.  Third and finally, there is another class of numerical methods\footnote{We include here the integral equation formulation of \citet{2012JETP..114..596P}, as it can be formally expressed as a classical algebraic eigenvalue problem.}, the ``matrix method'', which is based on solution of the linearised collisionless Boltzmann equation \citep{
1981SvA....25..533P,
1988MNRAS.231..935P,
1991MNRAS.248..494S,
1994ApJ...434...94B}, and a gentle didactic introduction can be found in \citet{2001ruag.conf..167W}. %
 For example, by use of such methods, much attention has been paid to the case of radially anisotropic systems, where the radial orbit instability leads to the presence of bars, which are of obvious interest from the point of view of the morphology of galaxies \citep[e.g.][]{1989ATsir1538....7P,
1990MNRAS.242..576A,
1995ApJ...440....5C,
1996ApJ...456..274C,
2006ApJ...637..717T,
2006ApJ...653...43M,
2007MNRAS.375.1157B,
2008ApJ...685..739B,
2009ApJ...704..372B,
2010MNRAS.405.2785M,
2015MNRAS.447...97G,
2017MNRAS.470.2190P}.

 But even to find that a system is stable is not the same as finding that it
 is dynamically inert.  In a clever extension of the matrix method \citet{1994ApJ...421..481W} found that certain stable spherical systems could possess perturbed, oscillatory modes that were weakly damped, on time scales much longer than the crossing time.  The systems he studied were the familiar models of \citet{1966AJ.....71...64K}, with central potential 
 $W_0 = 3,5,6$ and $7$.  These are all stable, by the theorems referred to above.

 While Weinberg's result is  interesting and surprising, it also seems to be important for understanding the collisional evolution of stellar systems.  This has been suspected from time to time over many years.  \citet{1996MNRAS.279..827T} showed that the rate of relaxation for loosely bound stars in a King model, as measured in an $N$-body system, was several times faster than the Chandrasekhar theoretical result, but in good agreement elsewhere\comment{, and included the influence of modes as one possible explanation.  The question was developed in detail by Weinberg himself \citep{1998MNRAS.297..101W}.}
Recently \citet{2018MNRAS.481.2041H}, who used a new formulation of relaxation theory which fully incorporates modes of oscillation such as those found by Weinberg, also found that the rate of relaxation for low binding energies considerably exceeded classical estimates in an isochrone model.  \comment{Finally, \citet{2019MNRAS.490..478L} have returned to the examination of this question with a very large set of short $N$-body models.}
 
In the present paper we focus on ``lop-sided'' (off-centre, \comment{or seiche}) modes, because these have the lowest damping rates of the modes which Weinberg discovered.  Such modes have long been of interest in flattened systems, especially spiral galaxies, but also in the dynamics of dark matter halos \citep{2009ApJ...697.2015S,
2011MSAIS..18..127D,
2011Ap.....54..184M,
2013MSAIS..25...55J,
2013ApJ...772..135Z,
2014MNRAS.439.1948Y,
2014MNRAS.444..352S,
2017A&A...600A..17P,
2017ApJ...844..130W,
2018ApJ...852...94V}.

Weinberg himself was able to demonstrate the existence of these modes
in a softened $N$-body model with $N = 10^4$ particles.  He used his
modal solution to construct the initial conditions, in such a way that
the initial peak density of the mode was 20\% of the background
equilibrium model.  He found that the decaying mode rotated with the
expected pattern speed.  He also observed similar coherent behaviour
even in models which were not carefully seeded, i.e. from initial
distributions drawn from the King distribution function.  Later, \citet{Purchase} used a self-consistent field code (mainly) to search for evidence of Weinberg's mode.  He seeded the system by transferring particles within the core (for example) from one side of the system to the other, creating a density perturbation qualitatively similar to that of an off-centre mode.  Though he summarised his results as ``inconclusive'' his caution was perhaps overstated.  

This \comment{summary of the literature} brings us to the present contribution, which aims to re-examine the occurrence of a weakly damped mode in an $N$-body system, but without softening, without any knowledge of the initial conditions (beyond guidance from Weinberg's results on the range of frequencies to examine),  with significant numbers of runs, \comment{and a duration exceeding several times the predicted period of Weinberg's mode}.  The set-up and analysis of the calculations is described in the following section, and the third section contains the results.   The final two sections discuss these and summarise   our conclusions.
 
\section{Methods}

\subsection{Computations}\label{sec:computations}

This paper focuses on the King model \citep{1966AJ.....71...64K} with concentration $W_0 = 5$.  The reasons for this choice were the following.  First, it was desired to choose a case in which the decay rate was nearly as small as possible, so that there was the best expectation of {\sl persistent} effects, which might be easier to detect from a long run of data.   According to Weinberg's results (his Table 1 and Fig.3) this meant $W_0 = 3$ or $5$.  Then it was desired that the real part of the eigenfrequency should be as large as possible, to optimise the prospects of detecting a (slowly decaying) periodic signal in the data; and of the two cases above, this implied a choice of $W_0 = 5$.  The initial conditions were sampled with a code written by us, and all particles have equal mass.

The \nbody simulations have been conducted with NBODY6 in a version adapted for use with a GPU \citep{2012MNRAS.424..545N}.   In this code the subroutine SCALE  recentres the initial conditions into a barycentric frame, a fact that will be of importance shortly (Sec.\ref{sec:analysis-spatial}).  The typical simulations reported here were run with $N=16384$ particles for a duration 
of 500 \hu\ \citep[HU;][]{1971Ap&SS..14..151H}, \comment{and data was output every time unit}.  The reason for this choice of duration was a compromise between two desiderata: first, to avoid a situation where the model evolves too much as a result of collisional relaxation (mainly, the development of core collapse); and, second, to give a sufficiently long data run to include several ``periods'' of the decaying mode found by Weinberg.  We consider these two issues now.

\begin{enumerate}
\item 
With the above initial conditions, the time of core collapse is found to be about 4000 HU.  Though the duration of our simulations might seem a high fraction of this time, the core collapse accelerates dramatically as it reaches completion; during the first 500 HU, however, the core radius (as determined by NBODY6) decreases typically from about  0.31 to about 0.25 HU.  For comparison, the King core radius changes by the same factor as $W_0$ increases from 5 to about 5.5.  From Weinberg's results (see above) one may estimate that the frequency of the mode changes by around 10\%, though the damping rate increases by about a factor 3.  Even so, the damping time scale will still be about twice the period.
\item
According to Weinberg, the frequency of the mode for $W_0 = 5$ is (in his units) 0.034.  His units are the same as H\'enon's units,  except that the virial radius is 2 instead of 1.  Therefore in HU the frequency is $0.034\times 2^{3/2}$, and the period is about 65 HU.  Therefore the duration of the computation is about $7.7$ times the period.  For comparison the initial predicted damping time scale is about 320 HU.
\end{enumerate}

\subsection{Analysis}

\subsubsection{Spatial structure}\label{sec:analysis-spatial}

In the matrix method used by Weinberg, the perturbed density is expanded in terms of surface harmonics denoted as $Y_{lm}(\theta,\phi)$, where $\theta,\phi$ are spherical polar angles, and $l,m$ are integers such that $\vert m\vert \le l$.  Though he considered both $l = 1$ and $l=2$, in this paper we focus on $l=1$.  Thus there are three surface harmonics, which we take to be $\sin\theta\cos\phi,\sin\theta\sin\phi,\cos\theta$.  These are just the angle-dependent parts of the coordinates $x,y,z$, and so we can think of the $l=1$ perturbation as a superposition of off-centre density distributions associated with the three coordinate axes.

As an example, consider the $x-$axis.  Then the density perturbation is of the form $\rho_x(r)\sin\theta\cos\phi$.  (Weinberg gives a graphical presentation of this mode in cross section on the $x,y$ plane in his Fig.4; see also Fig.\ref{fig:contour} in the present paper.)  Because of the real part of the frequency, the maximum of the density perturbation alternates on the positive and negative sides of the $x-$axis.  As Weinberg notes, one effect of this is that the ``density centre'', which is calculated by the $N$-body code, oscillates with the mode.  It is therefore one of the important sources of data on $l=1$ modes.

Incidentally, such oscillation along a single axis would be a very special manifestation of the mode.  Two modes, one each along the $x$- and $y$-axes, and $90^\circ$ out of phase, would give rise to a density centre rotating about the origin.  This was the kind of motion which Weinberg himself observed in a specially seeded $N$-body simulation.

  The density centre moves by only about 0.01 during our simulations (which are unseeded), whereas Weinberg shows that the structure of the mode extends to at least 0.5 HU.  Therefore some method of investigating the $N$-body data on much larger length scales than that of the density centre is desirable.  Now the density perturbation of interest may be written as
  \begin{equation}
    \rho_1(r,\theta,\phi) = a_1(r)\sin\theta\cos\phi + a_2(r)\sin\theta\sin\phi + a_3(r) \cos\theta.\label{eq:rho1}
  \end{equation}
  By orthogonality,
  \begin{equation}
    a_1(r) = \frac{3}{4\pi}\iint\rho(r,\theta,\phi)\sin\theta\cos\phi\sin\theta d\theta d\phi,
  \end{equation}
  with similar expressions for the other two amplitudes $a_2,a_3$, with the same normalisation constant.  In an $N$-body simulation this may be estimated in a spherical shell of volume $V$ by treating $\rho$ as a sum of delta functions, giving the estimate
  \begin{equation}
    \tilde{a}_1 = m\frac{3}{V}\sum\sin\theta\cos\phi,\label{eq:a1bar}
  \end{equation}
  where $m$ is the particle mass, and $ \tilde{a}_1$ can be thought of
  as an estimate of $a_1$ at some weighted mean radius inside the
  shell.  Similar expressions are used for $ \tilde{a}_2$ and $
    \tilde{a}_3$.  In our analysis six shells were used, separated by the five radii 0.2, 0.4, 0.6, 0.8 and 1.

\begin{figure}
  	\includegraphics[width=\columnwidth,trim={40 0 50 00}, clip=true]{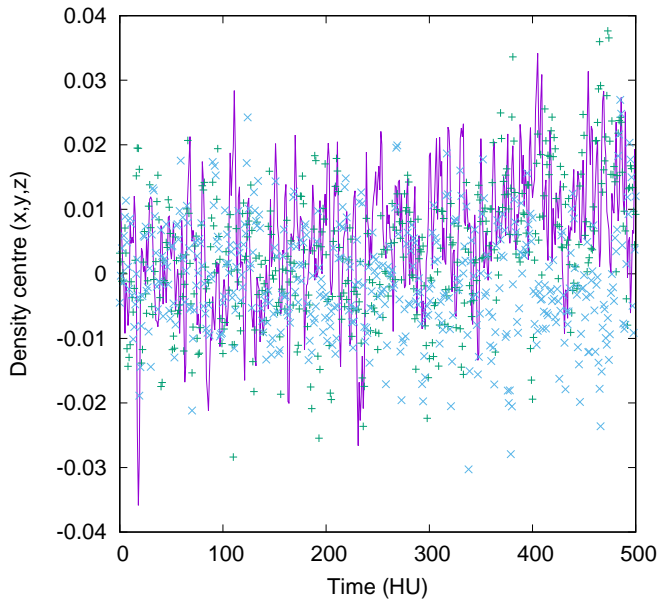}
    \caption{Evolution of the density centre in a rest-frame of the barycentre: solid ($x$), plusses ($y$), crosses ($z$).  \comment{The sampling interval is 1 time unit.}  Though $x$ shows a systematic drift, the overall drift is about 0.01 HU, which is of the order of the fluctuations in the three coordinates.  \comment{The probable reason for this drift is discussed in the text (Sec.\ref{sec:analysis-spatial}).}
    }
    \label{fig:dcxyz}
\end{figure}

    The above expressions presume an origin for the system of spherical coordinates.  Use of the density centre would be inappropriate for, as already mentioned, it partakes of the oscillations which we seek to observe, and its use would tend to suppress any signal of the oscillations, at least in the central parts of the system.  Therefore we adopt a point which is fixed in the original barycentric frame of the initial conditions.  
    \comment{There is then a danger that the barycentre of the bound system moves by recoil caused by escapers.  In the early stages of evolution (long before core collapse), results from \citet[][Fig.13]{2002MNRAS.336.1069B} show that a few escapers will appear with typical escape energy about 1/10 of the mean kinetic energy per particle in the system.  In our simulations, such an escaper could cause a displacement of the density centre of order 0.01 HU by the end of a simulation.  This is a plausible explanation of the drift (in the $x$-coordinate of the density centre) which is shown by the solid line in Fig.\ref{fig:dcxyz}.  But we have no suspicion that such a drift has any bearing on the significant results and conclusions in this paper.}
      
  Our choice of origin is actually the location of the density centre in the initial conditions, but results are not expected to be sensitive to the particular choice, provided that the origin is fixed in the barycentric frame and close to the barycentre.

\subsubsection{Temporal structure}\label{sec:temporal-results}

  The analysis considered so far is concerned with the spatial structure of the modes in an  $N$-body model.  Now we consider their time scales, in particular the period (i.e. $2\pi/\omega_r$, where $\omega_r$ is the real part of the eigenfrequency).  We expect (see Section \ref{sec:discussion}) that the modes are continually excited by the motions of the particles, and are constantly damped.  Two possible methods suggest themselves: autocorrelation analysis, and the study of power spectra.  \comment{These are, however, equivalent, as the power spectrum is basically the Fourier Transform of the autocorrelation.}

The autocorrelation method was used by \citet{1996MNRAS.282...19S} in a study of fluctuations in the position of the density centre in a 10000-body simulation starting from a Plummer model.  The density centre also features prominently in the work of the present paper, and their approach and results are obviously of interest.  Their analysis extends from about $t = 1200$ HU for a time interval of about 1200 HU until a point close to the time of core collapse (2380 HU).  Because of the considerable motion of the density centre, the data was detrended by applying a low pass filter.  Computation of the autocorrelation then yielded a function with a striking dominant periodicity with period about 40 HU\footnote{Their power spectrum is shown as Fig.9 in \citet{1996MNRAS.282...19S}, but we believe that the abscissa is incorrectly labelled as the period, whereas it is in fact the frequency (analogous to $k$ in our eq.(\ref{eq:fourier})).}.  Nevertheless, in the present paper we have not relied on this approach, though one result is given in 
Fig.\ref{fig:auto} and the associated discussion in Sec.\ref{sec:temporal-structure-results}.  Even there we have avoided the use of a low-pass filter, as a precaution against the possibility of introducing spurious periodicity.

In the present paper we adopt the methods of power spectral analysis. 
To be definite, we define the discrete Fourier transform of the $x$-coordinate of the density centre as
  \begin{equation}
    \bar{x}(k) = \frac{1}{500}\sum_{j=1}^{500}x_je^{-2\pi ijk/500},\label{eq:fourier}
  \end{equation}
where \comment{$x_j$ is the value at time $j$,} $i^2 = -1$, $k$ is a positive integer, and the power is 
  \begin{equation}
P(k) = \vert{\bar{x}(k)}\vert^2.    \label{eq:power}
  \end{equation}
The number 500 here refers to the specific situation of our analysis, where data were sampled at unit intervals of time to $t = 500$. \comment{ Often we add results from all three coordinates of the density centre, resulting in what we refer to as the {\sl total power} for a model.}  

Numerical estimation of the imaginary part of the frequency, i.e. the damping rate of a mode, depends in principle on the spectrum of the noise which excites the mode.  Consider, for example, the power spectrum of a damped simple harmonic oscillator excited by thermal noise, for which the equation of motion is
\begin{equation}
  \ddot x + \beta\dot x + \omega_0^2x = F(t).
\end{equation}
Then it can be shown\footnote{In this paragraph $k$ has the usual meaning in the context of the Fourier Integral Transform, which differs from that in eq.(\ref{eq:fourier}) by a factor $2\pi/500$.} that the power spectral density is given by
\begin{equation}
  P(k) \propto \frac{1}{(\omega_0^2 - k^2)^2 + \beta^2k^2},\label{eq:psd}
\end{equation}
where the constant of proportionality depends on the characteristics of the noise.  Evidently the maximum occurs at $k = \sqrt{\omega_0^2 - \beta^2/2} = \sqrt{\omega_r^2 - \omega_i^2}$, where the frequency of the unforced oscillator is written as $\omega_r + i\omega_i$.  If the oscillation is lightly damped, as predicted by Weinberg, this value of $k$ is close to $\omega_r$; in fact $k = \omega_r(1 + O(\omega_i^2/\omega_r^2))$.  Similarly, the full width at half maximum is 
\begin{equation}
\Delta k = 2\omega_i(1 + O(\omega_i^2/\omega_r^2)).  \label{eq:psd-width}
\end{equation}
Thus the width of the peak can, in principle, provide an estimate of the damping rate.

  \section{Results}\label{sec:results}

  \subsection{Spatial structure}\label{sec:structure}

  It is natural to expect that motion of the density centre mirrors that of any $l=1$ mode, especially near the density centre itself.  In fact Weinberg gives a relation $r_{shift}/r_c \simeq 0.4\varepsilon$ for the shift of the density centre, $r_{shift}$, in terms of the core radius $r_c$ and the ratio $\varepsilon \ltorder 0.3$ of the peak perturbed density to the background density there.  The numerical coefficient does not vary much with concentration $W_0$ over the range which he studied.  In our units $r_c\simeq0.4$.

\begin{figure}
  	\includegraphics[width=\columnwidth,trim={40 0 50 00}, clip=true]{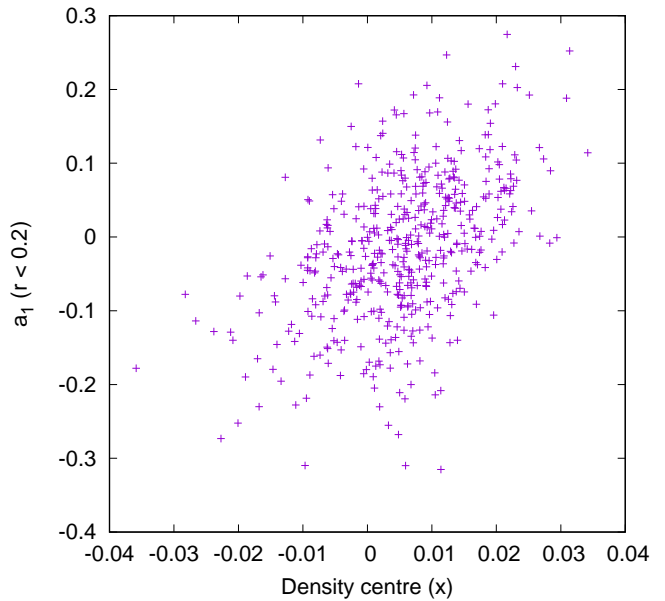}
    \caption{Scatter plot of the $x$-coordinate of the density centre against the estimate (eq.(\ref{eq:a1bar})) of the corresponding surface harmonic amplitude in the innermost shell, $r < 0.2$.  \comment{Each of the 500 outputs in the run gives one point.}
}
    \label{fig:xdca1bar}
\end{figure}

Fig.\ref{fig:xdca1bar} shows a closely related result from one $N$-body simulation with $N = 16384$.  There is a clear but noisy correlation between the $x$-coordinate of the density centre and the estimate of $a_1$ in the shell $r<0.2$, which is a measure of the density of an $l=1$ perturbation with maximum on the $x$-axis.  Very similar results are obtained for the corresponding pairs of variables $y, a_2$ and $z,a_3$ (see the first row of data in  Table \ref{tab:correlation}, discussed further below).

\begin{figure}
	\includegraphics[width=\columnwidth,trim={40 0 40 00}, clip=true]{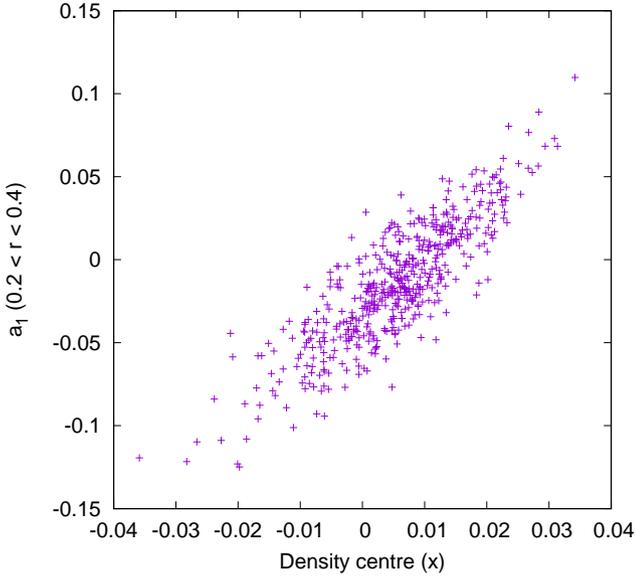}
    \caption{As Fig. \ref{fig:xdca1bar} but at radii $0.2<r<0.4$, i.e. the outer half of the core.
}
    \label{fig:xdca1bar-outercore}
\end{figure}

These results consider the density well inside the core radius ($r\ltorder 0.5r_c$), but Fig \ref{fig:xdca1bar-outercore} shows that the correlation strengthens in the outer core ($0.2<r<0.4$), in the sense that the relative scatter decreases.  At still larger radii the correlation weakens again, and outside the virial radius ($r>1$) it becomes a (weak) anticorrelation.

To make these statements quantitative, we give in Table \ref{tab:correlation} information obtained by linear regression in scatterplots such as those shown.  Figs. \ref{fig:xdca1bar} and \ref{fig:xdca1bar-outercore} correspond to the first two data entries in col.2, which gives the slopes of the regression fits along with the asymptotic 1-$\sigma$ confidence interval.  The increase in strength of the correlation is reflected in the decreasing  relative size of the confidence interval (from the first line to the second).  Besides the decreasing strength of the correlation at still larger radii, the change of sign of the slope at the largest radii ($r>1$) is also evident.  All these characteristics are clearly seen in the other two coordinates, i.e. the correlations between $y$ and $a_2$ (col.3) and those between $z$ and $a_3$ (col.4).

The negative slope at the largest radii is best constrained in $z$, and the reason for this may be linked with the larger trends in the $x-$ and $y-$coordinates of the density centre in this model (see Fig.\ref{fig:dcxyz}, and the last row of Table \ref{tab:correlation}).  If these are fitted with a linear function of $t$ and removed, the strength of the correlation (between $\tilde{a}_{i}$ and $x_{i}, i = 1,2$) is improved.  However, we have not used detrended data \comment{for the purpose of Table \ref{tab:correlation}.}

The data in the table is concerned with three components of the perturbed density, which fluctuate, but the strength of the density fluctuations can be characterised by multiplying the slope of the regression line by the rms of the coordinates 
of the density centre.  The result is an estimate of the typical density (but not its standard deviation, as it is evident that the sign at the largest radial bin is opposite that at the first five bins).  The estimate is given in Table \ref{tab:density}, which is the basis of discussion in Sec.\ref{sec:construction}; see also Fig.\ref{fig:contour}.

\begin{table}
	\centering
	\caption{The slope of the correlation between a component of the density perturbation and the corresponding coordinate of the density centre.}
	\label{tab:correlation}
	\begin{tabular}{lccr} 
		\hline
		Range of $r$ & $\tilde{a}_1/x$ & $\tilde{a}_2/y$ & $\tilde{a}_3/z$\\
		\hline
$(0,0.2)$&$ 4.5           (0.4)$&$ 4.8           (0.4)$&$ 5.1           (0.5)$\\
$(0.2,0.4)$&$ 3.02           (0.08)$&$ 3.20           (0.07)$&$ 3.17           (0.10)$\\
$(0.4,0.6)$&$ 1.01           (0.04)$&$ 1.08           (0.04)$&$ 1.09           (0.04)$\\
$(0.6,0.8)$&$ 0.33          (0.02)$&$ 0.32          (0.02)$&$ 0.32          (0.02)$\\
$(0.8,1)$&$ 0.088         (0.010)$&$ 0.094         (0.010)$&$ 0.047         (0.012)$\\
                $(1,\infty)$&$ -0.036        (0.010)$&$ -0.045        (0.010)$&$ -0.125         (0.008)$\\
                		\hline
                $\sigma$(dc)&0.0103&0.0105&0.0086\\

	\end{tabular}
{\raggedright  \comment{ Note: in the last three columns the figure in brackets is the\par standard deviation of the estimate.  
    \par}}
\end{table}

\begin{table}
	\centering
	\caption{Estimate of the typical density perturbation.  The tabulated value is an estimate of the value of $a_i(r)$ (see eq.(\ref{eq:rho1})) when the fluctuation in the $i^{th}$-coordinate of the density centre takes its 1-$\sigma$ value. A weighted average over all three coordinates is given.  The relative error is similar to that of the entries in Table \ref{tab:correlation}.}
	\label{tab:density}
	\begin{tabular}{lc} 
		\hline
		Range of radii & $\tilde{a}_i$\\
		\hline
$r<0.2$&0.047\\
$0.2<r<0.4$&0.031\\
$0.4<r<0.6$&0.011\\
$0.6<r<0.8$&0.0032\\
$0.8<r<1$&0.0008\\
$r>1$&-0.0007\\
		\hline
	\end{tabular}
\end{table}



\subsection{Temporal structure}\label{sec:temporal-structure-results}

In this subsection we turn to the time-dependence of the structures analysed in Sec.\ref{sec:structure}, \comment{with emphasis on} 
  power spectrum analysis, based on eqs.(\ref{eq:fourier}) and (\ref{eq:power}).  In view of the correlation reported in Table 	\ref{tab:correlation}, in principle this could be applied to either the coordinates of the density centre or the values of $a_i$ in the six radial bins that have been adopted.  It would also be possible to combine the latter by forming a weighted sum, using the results of Tab. \ref{tab:density}.  But the visual impression of the time series, even in a case where the correlation is strongest (e.g. Fig.\ref{fig:xdca1bar-outercore}), suggests that the temporal structures, \comment{though similar,} are less pronounced in the series for $\tilde{a}_i$ than in those for the density centre, and so we confine attention to this now.  \footnote{In this section, each time series of the density centre was first
detrended by fitting, and then subtracting, a linear function of
t. (Note added after publication.)}

\begin{figure}
	\includegraphics[width=\columnwidth,trim={30 0 50 00}, clip=true]{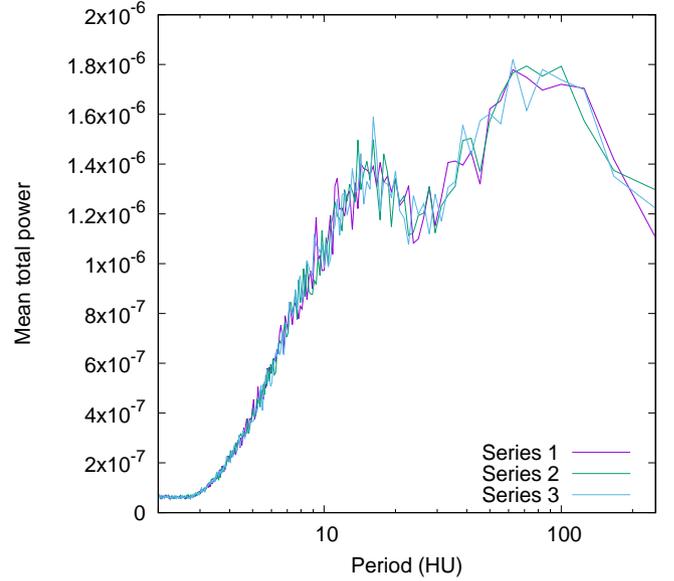}
    \caption{Average total power spectrum in three independent series of models, each series containing 100 simulations.
}
    \label{fig:sum2-ps}
\end{figure}

\comment{The power spectrum of a single model, even if the results for all three coordinates are combined, are at best suggestive, but far from persuasive.  Therefore we have carried out three sets of simulations, each consisting of 100 independent runs, and averaged the total power in all runs in each series.  Each series took about 4 days.  The result is shown in Fig.\ref{fig:sum2-ps}, which exhibits the average total power in each of the three series, in order to give the reader a feel for the significance of individual features.  Indeed only two convincing features are apparent: a broad peak around a maximum in the period range 60--70 (judged conservatively by the locations of the maxima across the three series), and a somewhat narrower peak around a maximum with period in the range 14--16. We consider each of these in turn.}

\begin{figure}
	\includegraphics[width=\columnwidth,trim={30 0 40 00}, clip=true]{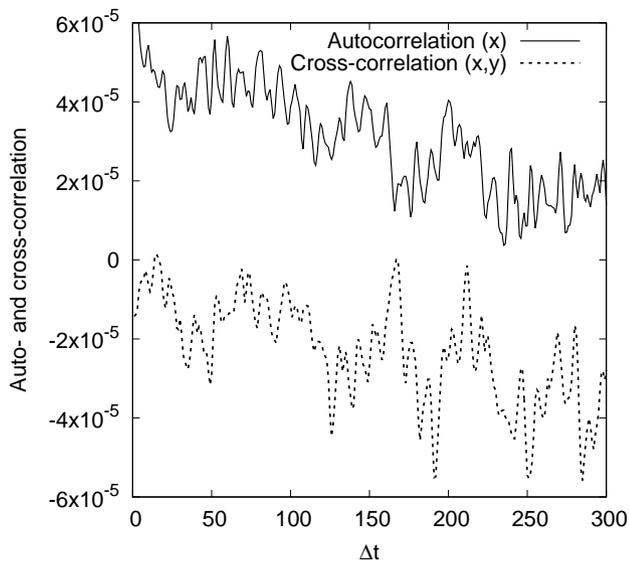}
    \caption{Autocorrelation of the $x$-coordinate of the density centre shown in Fig.\ref{fig:dcxyz}\comment{; and cross-correlation of $x,y$ in the same model, shifted vertically downwards by 0.00005 units}.  The  \comment{autocorrelation} is defined as $\sum x(t)x(t+\Delta t)/n$, where the interval between successive times is 1 HU, and $n$ is the number of terms in which $0\le t\le t+\Delta t \le 500$.  For larger $\Delta t$ (beyond those shown) the results become quite noisy.  \comment{The range of the ordinate has been chosen to reveal the features referred to in the text (Sec.\ref{sec:temporal-structure-results}), but also excludes an initial sharp drop in the autocorrelation (within the first 1 or 2 times), which is broadly similar to that shown by \citet{2019MNRAS.490..478L} in the autocorrelation of the position of the ``potential centre''.}}
    \label{fig:auto}
\end{figure}

\comment{The position of the feature with the longer range of periods is consistent with Weinberg's prediction (which corresponds to a period of approximately 65 in H\'enon units), and so we identify it with his prediction.  Further considerations of this feature, especially its width, are postponed to the discussion in Sec.\ref{sec:temporal-discussion}.  At the time of his work on the 1994 paper, Weinberg  (pers. comm.) noted signs of more strongly damped modes at higher frequency in some models, beyond the slowly damped modes like the one on which we have focused so far.  Our best interpretation of the short-period peak in Fig.\ref{fig:sum2-ps} is just such a mode, though there is nothing to justify any inference about its decay rate.

A rather striking impression of the two modes, especially the short-period one, is furnished by}
 the autocorrelation of the density centre, \comment{even if we confine attention to a single coordinate in a single model.}  
 An example is shown in Fig.\ref{fig:auto}.  There is a quite striking oscillation with an approximate period of about 10 HU (in the lag $\Delta t$).  \comment{Closer inspection, however, e.g. of the interval between successive extrema, reveals that the ``period'' itself fluctuates.  While this behaviour may help one to understand the breadth of the short-period peak in Fig.\ref{fig:sum2-ps},} 
     it comes no closer to providing a physical interpretation.  Incidentally, one may also persuade oneself that oscillations corresponding to Weinberg's mode are visible, but the power spectra are 
     more convincing.   \comment{Similar remarks may be made about a typical cross-correlation, which is also illustrated for the same model.  
     }

\section{Discussion}\label{sec:discussion}

\subsection{Spatial structure}\label{sec:construction}

\begin{figure}
    	\includegraphics[width=\columnwidth,trim={50 0 80 40},clip=true]{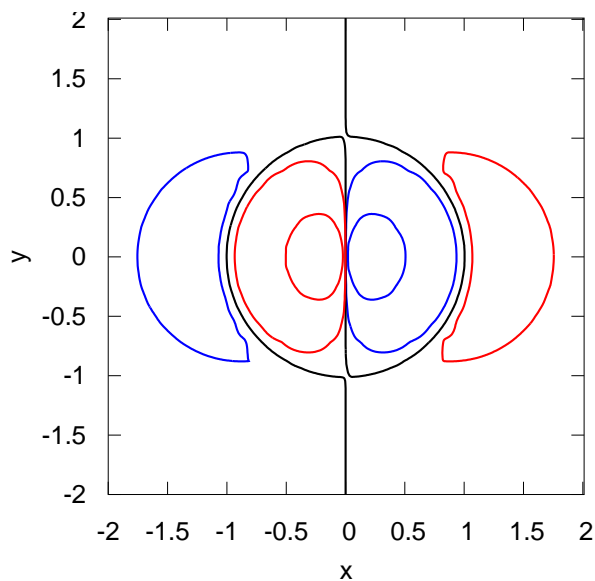}
    \caption{Contour diagram on the $x,y$ plane of the density perturbation $\rho_1 = a_1(r)\sin\theta\cos\phi$.  The method of construction is described in the text.  The contour levels are set at $-0.01$ and $-0.0005$ (blue, or darker grey), 0 (black) and 0.0005 and 0.01 (red, or lighter grey).  The irregularities in the large contours at right and left are artefacts of the plotting package, as are the gaps in the black contour near $(0,\pm1)$: the zero contour consists of two components, which are the $y$-axis, and a circle close to $r = 1$, respectively; the gaps at the ends of the $y$-diameter are incorrect.  
}
    \label{fig:contour}
\end{figure}

It is interesting to construct a contour map of the density perturbation, based on the data of Tab.\ref{tab:density}.  It is assumed that these data are values of the coefficient $a_1(r)$ at values of $r$ which are the midpoints of the ranges listed in the first column of the Table.  The last bin extends to infinity but the same notional bin-width is adopted, giving the value at $r = 1.1$, and for $r>1.1$ the functional form of $a_1(r)$ is assumed to be inversely proportional to $r$, \comment{merely for the sake of illustration}.  We also require that $a_1(0) = 0$, in order that the perturbed density is continuous at the origin.  From there to $r = 1.1$ piecewise linear interpolation is used.  We consider only perturbations which have the form of the first term on the right of eq.(\ref{eq:rho1}), and show a contour diagram of the perturbed density in the $x,y$ plane $\theta = \pi/2$ in Fig.\ref{fig:contour}.

As mentioned in Sec.\ref{sec:structure}, the values in the Table, and so the contour values in the figure, are merely typical.  Also, the signs may be reversed \comment{as the mode oscillates}.  Furthermore, in an actual $N$-body model, there will also be density perturbations corresponding to the other two terms in eq.(\ref{eq:rho1}), each with its own amplitude.  Also, these amplitudes vary with time.

The most interesting use of this figure is to compare it with a comparable figure constructed by Weinberg using the matrix method (his Fig.4).  The qualitative resemblance is striking.  His zero contour is not drawn, but one infers from his figure that the circular part must be close to a circle of radius 2, which corresponds to radius 1 in our units.  The maximum lies at about $(-0.6,0)$, or $(-0.3,0)$ in our units; in our diagram, however, the $x$-coordinate of the maximum appears to be closer to $x = -0.2$.  The main qualitative difference  between the two figures is the absence, in Weinberg's plot, of any contour outside the circular contour\footnote{\comment{The referee told us that a relevant figure  was omitted simply because he had been asked to save space.}}.  In Fig.\ref{fig:contour} this has the opposite sign to the value inside the circle, but on the same side of the origin.  \comment{(\citet{2019MNRAS.490..478L}, in their Fig.6, give a nice illustration which makes the same point.)}

This feature is important, because it relates to the fact that the centre of mass of the density perturbation must lie at the origin.  It is easy to see that \comment{its contribution to} the $x$-coordinate of the centre of mass is given by
\begin{equation}
  \bar x =  \frac{4\pi}{3}
  \int_0^\infty r^3 a_1(r) dr,
\end{equation}
in H\'enon units.
It is impossible to \comment{evaluate} 
this with the data in Tab.\ref{tab:density}, if only because the outer bin extends to infinity.  But if the integral is approximated by the midpoint rule, using the values in Table \ref{tab:density} at the six radii $0.1 (0.2) 1.1$, the final negative contribution is, by any reasonable estimate, too small compared with the sum of the five positive intervals, by a factor of a few.  A plausible reason for this finding is given in the following subsection, which re-examines  the interpretation of Fig.\ref{fig:contour} after first reviewing the power spectra.

\subsection{Temporal structure}\label{sec:temporal-discussion}

While Fig.\ref{fig:sum2-ps} gives evidence 
of motions \comment{at long periods similar to that of} 
Weinberg's slowly decaying mode, 
little can be said 
about \comment{their} 
observed decay rate.  \comment{If the broad width of the corresponding peak is taken at face value, and interpreted in terms of eq.(\ref{eq:psd}), one would conclude that the decay rate is roughly comparable with the frequency, which seems inconsistent } 
with Weinberg's prediction ($\omega_i/{\omega_r}\simeq 0.03$).  
\comment{It may very well be that decay is enhanced by two-body relaxation in what are relatively small simulations.  Equally, however, we might suppose that a coherent (slowly decaying) mode cannot be expected to occur if the motions are constantly and erratically reinforced by particle motions.  Similar remarks may be made about the short-period peak in the power spectrum of Fig.\ref{fig:sum2-ps}, though it seems relatively narrower than the long-period peak.}

\comment{The existence of at least two modes of oscillation deepens the interpretation of the density map we have constructed (Fig.\ref{fig:contour}).}  
Whatever the interpretation of the \comment{two} 
peaks \comment{in the power spectrum}, 
it cannot be claimed that the density map of Fig.\ref{fig:contour} is a map of \comment{a single mode, such as} Weinberg's mode.  It seems likely that, at small radii, all these motions correlate with the density centre, as we have found; whereas at large radii the anticorrelations (such as the negative values in Table \ref{tab:correlation}) may occur at different radii in different modes.  This may be one reason why the inferred density distribution apparently does not very well satisfy the requirement of a static  centre of mass (Sec.\ref{sec:construction}).

Before leaving the interpretation of the power spectra, it is worth noting that an interpretation in terms of \comment{conventional} modes may be misguided.  The simplest perturbation problem in collisionless stellar dynamics, i.e. Landau damping, can be solved in terms of a complete system of modes, but they have singular distribution functions (see, for example, \citealt[][Sec.5.2.4]{2008gady.book.....B}).  Whether such an idea can develop into a successful understanding of the observed power spectra in King's model is not, however, something that will be considered in this paper.

\subsection{Future developments}

While extension into the realm of singular modes seems to be beyond the power of current methods, there are several interesting ways in which the experiments reported in this paper can be developed.  In the first instance, though a few calculations have been carried out  with other models, very little has been done to explore the following issues.

\begin{enumerate}
\item $N$-dependence.  It is assumed that the motions that have been observed are collisionless phenomena.  Therefore the $N$-dependence of the time scales\comment{, as encapsulated in the power spectrum,} should be trivial, but that remains to be verified.
\item $W_0$-dependence.  Now that methods for \comment{searching for and analysing modes like} 
  Weinberg's mode have been described, it may be a straightforward matter to extend the study to other models in King's sequence.  Weinberg's paper gives results for a few other values of the scaled central potential, $W_0$, and can act as a guide for further numerical experiments.  But different $W_0$ have different frequencies and decay rates, and different core collapse times.  Therefore the compromise that needs to be struck between a simulation that is long enough by comparison with the \comment{period} 
  and short enough by comparison with the core collapse time needs to be borne in mind.
\item Model-dependence and anisotropy.  A King model with $W_0 =5$ has a concentration, measured by the ratio of the core and virial radii, of about 0.407, very similar to that of a Plummer model (concentration $3\pi/(16\sqrt{2})\simeq 0.4165$).  This suggests that the existence and frequency  of slowly damped modes may be similar.  Nevertheless it must be noted that the Plummer model is of infinite extent, unlike the King model (which is one reason why Weinberg, using the matrix method, chose it.)   In principle the existence of the finite edge in the King model might play an essential role in its modal structure, despite the low density and low modal amplitude near the edge.  
Still, the modal structure of the Plummer model is of obvious interest, not just because of the universal appeal of this model, but also because of the existence of a remarkable series of anisotropic models of which it is a member \citep{1987MNRAS.224...13D}; these are easy to generate with existing, public software by P. Breen\footnote{{It is available at https://github.com/pgbreen/PlummerPlus .}} \citep{L2}.
  \item Rotation.  It is easy to set a stellar system in rotation \citep{1960MNRAS.120..204L,L2}, and it would be interesting to observe the effect on \comment{results of this paper.} 
    Indeed this was the lead author's motivation in taking up the problem described in this paper, in the hope that rotation might destabilise the mode.  Here he was guided by an old, simple dynamical model by \citet{Lamb1908}, on the stability of a particle moving inside a rough, rotating, spherical bowl.  This motivation may sound naive, but the example was used by \citet[][Sec.185]{1928asco.book.....J} as an introductory example to the stability of rotating fluid masses.
\end{enumerate}

Another important extension of this research is theoretical.  Enough has been said to show that the motion of the density centre is not simply the kind of fluctuation that would occur in the motion of $N$ particles in a smooth potential; it is part of 
collective motion{s} involving the entire self-gravitating system.  Nevertheless, one might suppose that the motions of the particles occasionally exhibit large-amplitude fluctuations which excite detectable modal behaviour, of the kind demonstrated in this paper.   A mathematical model of this excitation process is, however, lacking, and without it the estimation of the $N$-dependence of the amplitude of the mode is a matter of guesswork.

\section{Conclusions}

This paper has considered self-excited collective oscillations (damped modes) of a spherical $N$-body system.  The methods are numerical, though they were intended to test a theoretical prediction by \citet{1994ApJ...421..481W}, which used the matrix method of stellar dynamical stability.  The modes studied are so-called ``$l = 1$'' modes, often referred to as lop-sided modes, and general motions of this form can be regarded as a superposition of motions aligned on the three coordinate axes.  By study of \comment{over 300} 
King models with concentration $W_0 = 5$ and $N = 16384$ equal-mass particles, it is shown in this paper that the motion of the density centre correlates with perturbations in the density throughout the system inside the virial radius (approximately), and anti-correlates with density fluctuations outside this radius.  From examination of power spectra, it is found that significant fluctuations in the density centre have a frequency within \comment{a range including} 
that predicted by Weinberg; this is observed in 
all \comment{large subsets of the data which have been} 
examined. 
High powers are also exhibited \comment{in a different range} 
of periods \comment{which peaks at a period of about five crossing times, which is smaller than that} 
of Weinberg's mode \comment{by a factor of about 4},   
but these motions have no theoretical interpretation \comment{as yet}.  Much work remains to be done in understanding how the motions are excited, and how they are altered by varying the parameters of the model; in particular concentration, isotropy and rotation.


\section*{Acknowledgements}

\comment{The authors are very grateful to Martin Weinberg for an engaged and instructive referee report, which resulted in a much increased computational sample and other substantial improvements.}
Support for DCH and PGB from
the Leverhulme Trust (RPG-2015-408) is gratefully acknowledged.   ALV thanks a JSPS International Fellowship and Grant-in-Aid (KAKENHI-18F18787) and the Institute for Astronomy at the University of Edinburgh.
We all thank Steve Law and David Marsh for tending the GPUs in the School of Mathematics, but some calculations were also conducted at the Edinburgh Compute and Data Facility.












\bsp	
\label{lastpage}
\end{document}